\documentclass[10pt]{article}
\usepackage{latexsym}
\usepackage{amssymb}
\usepackage{amsmath}
\usepackage{amscd}
\usepackage{amsthm}
\usepackage[left=2cm,top=2.5cm,right=2.5cm,bottom=1.5cm]{geometry}
\usepackage[dvips]{graphicx}
\usepackage{hyperref}
\begin{document}
\begin{center}
\large{\bf{Two-fluid atmosphere from decelerating to accelerating FRW dark energy models}} \\
\vspace{10mm}
\normalsize{Anirudh Pradhan}  \\
\vspace{5mm} 
\normalsize{Department of Mathematics, Hindu Post-graduate College, Zamania-232 331, Ghazipur, India \\
E-mail : pradhan@iucaa.ernet.in, pradhan.anirudh@gmail.com} \\
\end{center}
\vspace{10mm}
%\date{}
%\maketitle
\begin{abstract}
The evolution of the dark energy parameter within the scope of a spatially homogeneous and isotropic 
Friedmann-Robertson-Walker (FRW) model filled with perfect fluid and dark energy components is studied 
by generalizing the recent results (Amirhashchi et al. in Int. J. Theor. Phys. 50: 3529, 2011b). The two 
sources are claimed to interact minimally so that their energy momentum tensors are conserved separately. 
The conception of time-dependent deceleration parameter (DP) with some suitable assumption yields an average 
scale factor $a = [\sinh (\alpha t)]^{\frac{1}{n}}$, with $\alpha$ and $n$ being positive arbitrary constants. 
For $0 < n \leq 1$, this generates a class of accelerating models while for $n > 1$, the models of universe 
exhibit phase transition from early decelerating phase to present accelerating phase which is supported with 
the results from recent astrophysical observations. It is observed that the transition red shift ($z_{t}$) for 
our derived model with $q_{0} = -0.73$  is $\cong 0.32$. This is in good agreement with the cosmological 
observations in the literature. Some physical and geometric properties of the model along with physical acceptability 
of cosmological solutions have been discussed in detail.
\end{abstract}
%%%%%%%%%%%%%%%%%%%%%%%%%%%%%%%%%%%%%%%%%%%%%%%%%%%%%%%%%%%%%%%%%%%%%%%%%%%%%%%%%%%%%%%%%%%%%%%%%%%%%
%%%%%%%%%%%%%%%%%%%%%%%%%%%%%%%%% SECTION 1 %%%%%%%%%%%%%%%%%%%%%%%%%%%%%%%%%%%%%%%%%%%%%%%%%%%%%%%%%%
\section{Introduction}
The 2011 Nobel Prize in Physics has been awarded to Perlmutter, Riess and Schmidt for their very careful 
and independent observations of distant supernovae, apparently showing the rate of expansion of the 
Universe is accelerating. In 1998, the promulgated observations of Type Ia supernova (SNeIa) established 
that our universe is presently accelerating (Perlmutter et al. 1998, 1999; Riess et al. 1998) and recent 
observations of SNeIa of high sureness level (Tonry et al. 2003; Riess et al. 2004; Clocchiatti et al. 2006) 
have further confirmed this. In addition, measurements of the cosmic microwave background (CMB) (Bennett 
et al. 2003) and large scale structure (LSS) (Tegmark et al. 2004a) strongly indicate that our universe is 
dominated by a component with negative pressure, dubbed as dark energy. These also indicate that the universe 
has a flat geometry on large scales. Because there is not enough matter in the universe $-$ ether ordinary 
or dark matter $-$ to produce this flatness, the difference must be attributed to a ``dark energy''. 
This same dark energy causes the acceleration of the expansion of the universe. In addition, the effect of dark 
energy seems to vary, with the expansion of the universe slowing down and speeding up over different time. The 
Wilkinson Microwave Anisotropy Probe (WMAP) satellite experiment suggests $73\%$ content of the universe in the 
form of dark energy, $23\%$ in the form of non-baryonic dark matter and the rest $4\%$ in the form of the usual 
baryonic matter as well as radiation. \\

High-precision measurements of expansion of the universe are required to understand how the expansion rate changes 
over time. In general relativity, the evolution of the expansion rate is parameterized by the cosmological equation 
of state (the relationship between temperature, pressure, and combined matter, energy, and vacuum energy density for 
any region of space). Measuring the equation of state for dark energy is one of the biggest efforts in observational 
cosmology today. The DE model has been characterized in a conventional manner by the equation of state (EoS) parameter 
$\omega^{de} = \frac{p^{de}}{\rho^{de}}$ which is not necessarily constant, where $\rho^{de}$ is the energy density 
and $p^{de}$ is the fluid pressure (Carroll and Hoffman 2003). The $\omega^{de}$ lies close to $-1$: it would be 
equal to $-1$ (standard $\Lambda$CDM cosmology), a little bit upper than $-1$ (the quintessence dark energy) or 
less than $-1$ (phantom dark energy). While the possibility $\omega \ll -1$ is ruled out by current 
cosmological data from SN Ia (Supernovae Legacy Survey, Gold sample of Hubble Space Telescope) (Riess et al. 2004; 
Astier et al. 2006), CMB (WMAP, BOOMERANGE) (Eisentein et al. 2005; MacTavish et al. 2006) and large scale structure 
(Sloan Digital Sky Survey) data (Komatsu et al. 2009), the dynamically evolving DE crossing the phantom divide line 
(PDL) ($\omega = -1$) is mildly favoured. The simplest candidate for the dark energy is a cosmological constant 
$\Lambda$, which has pressure $p^{(de)} = - \rho^{(de)}$. Specifically, a reliable model should explain why the 
present amount of the dark energy is so small compared with the fundamental scale (fine-tuning problem) and why 
it is comparable with the critical density today (coincidence problem) (Copeland et al. 2006). That is why, the 
different forms of dynamically changing DE with an effective equation of state (EoS), 
$\omega^{(de)} = p^{(de)}/\rho^{(de)} < -1/3$, have been proposed in the literature. Some other limits obtained from 
observational results coming from SNe Ia data (Knop et al. 2003) and combination of SNe Ia data with CMBR anisotropy and 
galaxy clustering statistics (Tegmark et al. 2004) are $-1.67 < \omega < -0.62$ and $-1.33 < \omega < - 0.79$, respectively. 
The latest results in 2009, obtained after a combination of cosmological datasets coming from CMB anisotropies, luminosity 
distances of high redshift type Ia supernovae and galaxy clustering, constrain the dark energy EoS to 
$-1.44 < \omega < -0.92$ at $68\%$ confidence level (Komatsu et al. 2009; Hinshaw et al. 2009). \\

Cai et. al. (2010) review all scalar-field based dark energy. Chen et al. (2009) perform a detailed phase-space analysis of 
various phantom cosmological models where dark energy sector interacts with the dark matter one. Jamil et al. (2010) discussed 
thermodynamics of dark energy interacting with dark matter and radiation. Recently, Singh and Chaubey (2012, 2013) obtained 
interacting two-fluid scenario for dark energy in anisotropic Bianchi type space-times. Reddy and Kumar (2013) discussed 
two-fluid scenario for dark energy model in a scalar tensor theory of gravitation. Naidu et al. (2012a, 2012b) and Reddy et al. 
(2012) discussed Bianchi type-V, III and five dimensional dark energy models respectively in scalar-tensor theory of gravitation. 
Recently, Amirhashchi et al. (2011a, 2011b), Pradhan et al. (2011), Saha et al. (2012) have studied an interacting and non-interacting 
two-fluid scenario for dark energy models in FRW universe. Kumar (2011) studied some FRW models of accelerating universe with dark 
energy. In this paper we study the evolution of the dark energy parameter within the framework of a FRW cosmological model filled 
with two fluids (barotropic and dark energy) by revisiting the recent work of Amirhashchi et al. (2011b) and obtained more general 
results. The cosmological implications of this two-fluid scenario will be discussed in detail in this paper. In doing so the two 
sources are claimed to interact minimally so that their energy momentum tensors are conserved separately. The out line of the paper 
is as follows: In Sect. $2$, the metric and the basic equations are described. Sections $3$ deals with the solutions of the field equations. 
The physical significances of the model s are discussed in Sect. $4$. Physical acceptability of the model is discussed in Sect. $5$. 
Finally, conclusions are summarized in the last Sect. $6$.  
%%%%%%%%%%%%%%%%%%%%%%%%%%%%%%%  SECTION 2  %%%%%%%%%%%%%%%%%%%%%%%%%%%%%%%%%%%%%%%%%%%%%%%%%%%%%%
\section{Metric and field  equations}
In standard spherical coordinates $(x^{i}) = (t, r, \theta, \phi)$, a spatially homogeneous and isotropic 
FRW line-element has the form (in units $c = 1$)
\begin{equation}
\label{eq1} ds^{2} = -dt^{2} + a^{2}(t)\left[\frac{dr^{2}}{1 - kr^{2}} + r^{2}\left(d\theta^{2} + 
\sin^{2}\theta d\phi^{2}\right)\right],
\end{equation}
where $a(t)$ is the cosmic scale factor, which describes how the distances (scales) change in an expanding 
or contracting universe, and is related to the redshift of the 3-space; $k$ is the curvature parameter, 
which describes the geometry of the spatial section of space-time with closed, flat and open universes 
corresponding to $k = -1$, $0$, $1$, respectively. The coordinates $r$, $\theta$ and $\phi$ in the metric 
(\ref{eq1}) are comoving coordinates. The FRW models have been remarkably successful in describing the 
observed nature of universe. \\

The Einstein's field equations in case of a mixture of perfect fluid and DE components, in the units 
$8\pi G = c =1$, read as
\begin{equation}
\label{eq2} R^{j}_{i} - \frac{1}{2}R g^{j}_{i} = T^{j}_{i},
\end{equation}
where $T^{j}_{i} = T_{i}^{(m)j} + T_{j}^{(de)j}$ is the overall energy momentum tensor with $T_{j}^{(m)j}$ and 
$T_{i}^{(de)j}$ as the energy momentum tensors of ordinary matter and DE, respectively. These are given by
\[
T_{i}^{(m)j} = {\rm diag}\left[-\rho^{(m)}, \;  p^{(m)}, \;  p^{(m)}, \; p^{(m)}
\right]
\]
\begin{equation}
\label{eq3}  = {\rm diag}\left[-1, \; \omega^{(m)}, \; \omega^{(m)}, \; \omega^{(m)}\right]\rho^{(m)} \; ,
\end{equation}
and
\[
T_{i}^{(de)j} = {\rm diag}\left[-\rho^{(de)}, \; p^{(de)}, \; p^{(de)}, \; p^{(de)}\right]
\]
\begin{equation}
\label{eq4}  = {\rm diag}\left[-1, \omega^{(de)}, \omega^{(de)}, \omega^{(de)}\right]\rho^{(de)} \; ,
\end{equation}
where $\rho^{(m)}$ and $p^{(m)}$ are, respectively the energy density and the isotropic pressure of the perfect fluid 
component or ordinary baryonic matter while $\omega^{(m)} = p^{(m)}/\rho^{(m)}$ is its EoS parameter. 
Similarly, $\rho^{(de)}$ and $p^{(de)}$ are, respectively the energy density and pressure of the DE component
while $\omega^{(de)} = p^{(de)}/\rho^{(de)}$ is the corresponding EoS parameter. \\

In a comoving coordinate system, the field equations (\ref{eq2}), for the FRW space-time (\ref{eq1}), with
(\ref{eq3}) and (\ref{eq4}), read as
\begin{equation}
\label{eq5} 2\frac{\ddot{a}}{a} + \frac{\dot{a}^{2}}{a^{2}} + \frac{k}{a^{2}} = -\omega^{(m)}\rho^{(m)} 
- \omega^{(de)}\rho^{(de)},
\end{equation}
\begin{equation}
\label{eq6} 3\left(\frac{\dot{a}^{2}}{a^{2}} + \frac{k}{a^{2}}\right) = \rho^{(m)} + \rho^{(de)}.
\end{equation}
Here the over dot denotes derivative with respect to $t$. \\

The law of energy conservation equation $T^{ij}_{~ ~ ;j} = 0$ yields
\begin{equation}
\label{eq7} \dot{\rho}^{(m)} + 3(1 +\omega^{(m)})\rho^{(m)}H + \dot{\rho}^{(de)} + 3(1 + \omega^{(de)})
\rho^{(de)}H = 0,
\end{equation}
where $H = \frac{\dot{a}}{a}$ is the Hubble parameter. 
%%%%%%%%%%%%%%%%%%% Figure 1 %%%%%%%%%%%%%%%%%%%%%%%%%%%%%%%%%%%%%%%%%%%%%%%%%%%%%%%%%%%%%%
\begin{figure}[ht]
\centering
\includegraphics[width=12cm,height=8cm,angle=0]{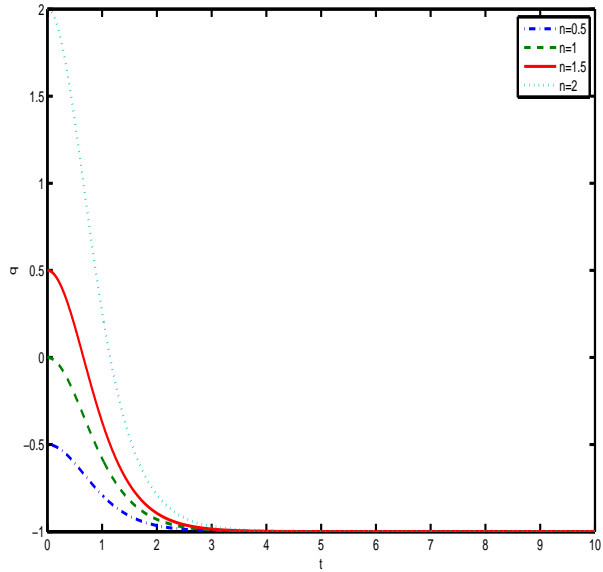} \\
\caption{The plot of deceleration parameter $q$ vs time $t$ for $\alpha = 1$.}
\end{figure}
%%%%%%%%%%%%%%%%%%%%%%%%%%%%%%%%%% %%%%%%%%%%%%%%%%%%%%%%%%%%%%%%%%%%%%%%%%%%%%%%%%%%%%%%%%%%%
%%%%%%%%%%%%%%%%%%% Figure 2 %%%%%%%%%%%%%%%%%%%%%%%%%%%%%%%%%%%%%%%%%%%%%%%%%%%%%%%%%%%%%%
\begin{figure}[ht]
\centering
\includegraphics[width=20cm,height=8cm,angle=0]{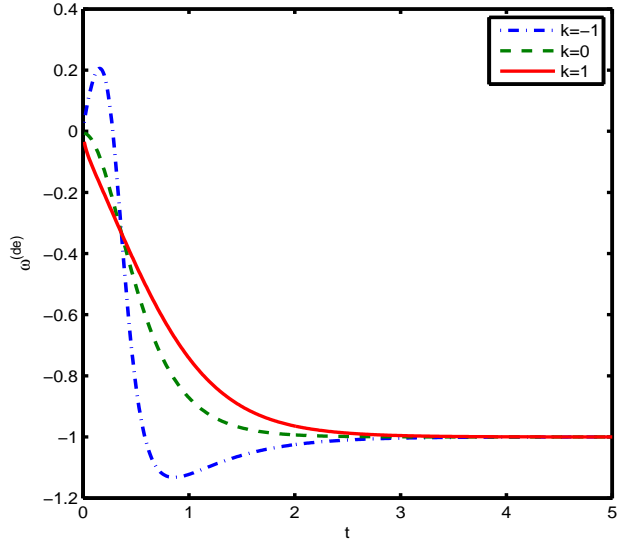} \\
\caption{The plot of DE EoS parameter $\omega^{de}$ versus $t$ for closed, flat and open universe
for $\omega^{m} = 0.01$, $\alpha = 1.5$, $n = 1.5$, $\rho_{0} = 1$.}
\end{figure}
%%%%%%%%%%%%%%%%%%%%%%%%%%%%%%%%%% %%%%%%%%%%%%%%%%%%%%%%%%%%%%%%%%%%%%%%%%%%%%%%%%%%%%%%%%%%%
%%%%%%%%%%%%%%%%%%%%%%%%%%%%%%%%%%%%%%%%%%%%%%%%%%%%%%%%%%%%%%%%%%%%%%%%%%%%%%%%%%%%%%%%%%%%%%%%%%%%
%%%%%%%%%%%%%%%%%%%%%%%%%%%%%%%  SECTION 3  %%%%%%%%%%%%%%%%%%%%%%%%%%%%%%%%%%%%%%%%%%%%%%%%%%%%%%%%%%%
\section{Solution of field  equations}
The field equations (\ref{eq5}) and (\ref{eq6}) involve five unknown variables, viz., $a$, $\omega^{(m)}$, 
$\omega^{(de)}$, $\rho^{(m)}$  and $\rho^{(de)}$. Therefore, to find a deterministic solution of the
equations, we need three suitable assumptions connecting the unknown variables. \\

In order to solve the field equations completely, we first assume that the perfect fluid and DE components interact 
minimally. Therefore, the energy momentum tensors of the two sources may be conserved separately. \\

Following Akarsu and Kilinc (2010a), first we assume that the perfect fluid and DE components interact 
minimally. Therefore, the energy momentum tensors of the two sources may be conserved separately. \\

The energy conservation equation ($T^{(m)ij}_{;j} = 0$) of the perfect fluid leads to
\begin{equation}
\label{eq8} \dot{\rho}^{(m)} + 3(1 + \omega^{(m)})\rho^{(m)}H = 0,
\end{equation}
whereas the energy conservation equation ($T^{(de)ij}_{;j} = 0$) of the DE component yields
\begin{equation}
\label{eq9} \dot{\rho}^{(de)} + 3(1 + \omega^{(de)})\rho^{(de)}H = 0.
\end{equation}
Following Akarsu and Kilinc (2010a,b,c), and Kumar and Yadav (2011), we assume that the EoS parameter of the 
perfect fluid to be a constant, that is,
\begin{equation}
\label{eq10} \omega^{(m)} = \frac{p^{(m)}}{\rho^{(m)}} = const.,
\end{equation}
while $\omega^{(de)}$ has been admitted to be a function of time since the current cosmological data from SN Ia, 
CMBR and large scale structures mildly favor dynamically evolving DE crossing the PDL as discussed in previous section. \\

Eq. (\ref{eq8}) can be integrated to lead
\begin{equation}
\label{eq11} \rho^{(m)} = \rho_{0}a^{-3(1+\omega^{(m)})},
\end{equation}
where $\rho_{0}$ is a positive constant of integration. \\

Firstly, we define the deceleration parameter q as
\begin{equation}
\label{eq12} q = -\frac{a\ddot{a}}{\dot{a}^{2}} = - \left(\frac{\dot{H} + H^{2}}{H^{2}}\right)
 = b(t) ~ ~\mbox{say}.
\end{equation}
The time-dependent behaviour of $q$ is supported by recent observations of SNe Ia (Riess et al., 1998; Perlmutter 
et al., 1998, 1999; Tonry et al., 2003; Clocchiatti et al., 2006) and CMB anisotropies (Bennett et al., 2003; de Bernardis 
et al., 2000; Hanany et al., 2000). These observations clearly indicate an accelerating expansionary universe at present 
which was decelerating in past. In their preliminary analysis, it was found that the SNe data favour recent acceleration 
($z < 0.5$) and past deceleration ($z > 0.5$). Recently, the High-Z Supernova Search (HZSNS) team have prevailed transition 
redshift $z_{t} = 0.46 \pm 0.13$ at ($1\; \sigma$) c.1. (Riess et al., 2004) which has been further amended to 
$z_{t} = 0.43 \pm 0.07$ at ($1\; \sigma$) c.1. (Riess et al., 2007). The Supernova Legacy Survey (SNLS) (Astier et al., 2006), 
as well as the one recently compiled by Davis et al. (2007), yields $z_{t} \sim 0.6 (1 \; \sigma)$ in better agreement with 
the flat $\Lambda$CDM model ($z_{t} = (2\Omega_{\Lambda}/\Omega_{m})^{\frac{1}{3}} - 1 \sim 0.66$). Thus, the DP which by definition 
is the rate with which the universe decelerates, must show signature flipping (see the Refs. Riess et al., 2001; Padmanabhan 
and Roychowdhury 2003; Amendola 2003) between positive and negative values. Following Pradhan and Otarod (2006), Akarsu and Dereli (2012),
Pradhan et al. (2012), Chawla et al. (2012), Chawla and Mishra (2013), Mishra et al. (2013), Pradhan et al. (2013), Pradhan (2013), 
Amirhashchi et al. (2013), we have discussed the model of the universe with variable DP. \\
%%%%%%%%%%%%%%%%%%% Figure 3 %%%%%%%%%%%%%%%%%%%%%%%%%%%%%%%%%%%%%%%%%%%%%%%%%%%%%%%%%%%%%%
\begin{figure}[ht]
\centering
\includegraphics[width=20cm,height=8cm,angle=0]{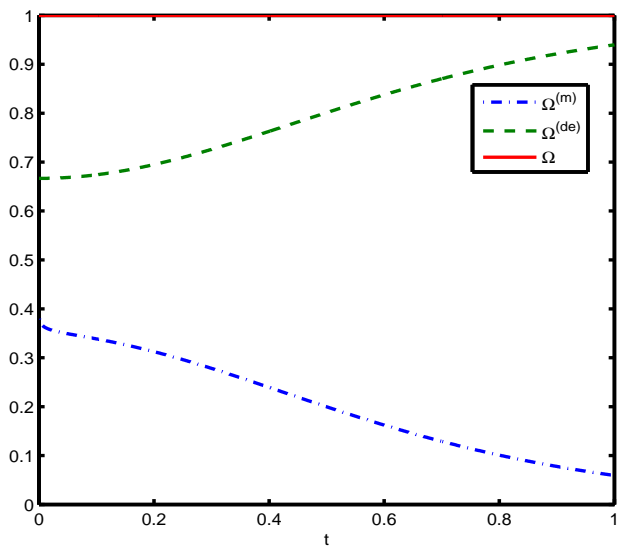} \\
\caption{The plot of $\Omega^{m}$, $\Omega^{de}$, $\Omega$ versus $t$ for flat universe ($k = 0$) 
for $\omega^{m} = 0.01$, $\alpha = 1.5$, $n = 1.5$, $\rho_{0} = 1$.}  
\end{figure}
%%%%%%%%%%%%%%%%%%%%%%%%%%%%%%%%%% %%%%%%%%%%%%%%%%%%%%%%%%%%%%%%%%%%%%%%%%%%%%%%%%%%%%%%%%%%%
Equation (\ref{eq12}) may be rewritten as
\begin{equation}
\label{eq13} \frac{\ddot{a}}{a} + b\frac{\dot{a}^{2}}{a^{2}} = 0.
\end{equation}
In order to solve the Eq. (\ref{eq16}), we assume $b = b(a)$. It is important to note here that one can assume
$b = b(t) = b(a(t))$, as $a$ is also a time dependent function. It can be done only if there is a one to one
correspondences between $t$ and $a$. But this is only possible when one avoid singularity like big bang or big
rip because both $t$ and $a$ are increasing functions.  \\

The general solution of Eq. (\ref{eq13}) with the assumption $b = b(a)$, is obtained as
\begin{equation}
\label{eq14} \int e^{\int\frac{b}{a}da}da = t+k,
\end{equation}
where $k$ is an integrating constant. \\

One cannot solve (\ref{eq14}) in general as $b$ is variable. So, in order to solve the problem completely,
we have to choose $\int\frac{b}{a}da$ in such a manner that (\ref{eq14}) be integrable without any loss
of generality. Hence we consider
\begin{equation}
\label{eq15} \int\frac{b}{a}da = \ln f(a),
\end{equation}
which does not affect the nature of generality of solution. Hence from (\ref{eq14}) and (\ref{eq15}),
we obtain
\begin{equation}
\label{eq16} \int f(a)da = t + k.
\end{equation}
Of course the choice of $f(a)$, in (\ref{eq16}), is quite arbitrary but, since we are looking for
physically viable models of the universe consistent with observations, we consider
\begin{equation}
\label{eq17} f(a) = \frac{na^{n-1}}{\alpha \sqrt{1+a^{2n}}},
\end{equation}
where $\alpha$ is an arbitrary constant and $n$ is a positive constant. In this case, on integrating Eq.
(\ref{eq16}) and neglecting the integration constant $k$, we obtain the exact solution as
\begin{equation}
\label{eq18} a(t) = (\sinh(\alpha t))^{\frac{1}{n}}.
\end{equation}
This relation (\ref{eq18}) generalizes the value of scale factor obtained by Amirhashchi et al. (2011b) 
and Pradhan et al. (2012) in connection with the study of dark energy models respectively in FRW 
and Bianchi type-$VI_{0}$ space-times. \\

From (\ref{eq18}), we obtain the time varying deceleration
parameter as
\begin{equation}
\label{eq19} q = -\frac{a\ddot{a}}{\dot{a}^{2}} = n\left[1 - (\tanh(\alpha t))^{2}\right]- 1.
\end{equation}
From Eq. (\ref{eq19}), we observe that $q > 0$ for $t < \frac{1} {\alpha}\tanh^{-1}(1 - \frac{1}{n})^{\frac{1}{2}}$ and $q < 0$
for $t > \frac{1}{\alpha}\tanh^{-1}(1 - \frac{1}{n})^{\frac{1}{2}}$. It is also observed that for $0 < n \leq 1$, our model is 
in accelerating phase but for $n > 1$, our model is evolving from decelerating phase to accelerating phase. Also, recent 
observations of SNe Ia, expose that the present universe is accelerating and the value of DP lies to some place in the range 
$-1 \leq q < 0$. It follows that in our derived model, one can choose the value of DP consistent with the observations. Figure $1$ 
depicts the variation of the deceleration parameter ($q$) versus time $(t$) which gives the behavior of $q$ for  different values 
of $n$. It is also clear from the figure that for $n \leq 1$, the model is evolving only in accelerating phase whereas for $n > 1$ 
the model is evolving from the early decelerated phase to the present accelerating phase. \\

Using (\ref{eq18}), the model (\ref{eq1}) becomes
\begin{equation}
\label{eq20}
ds^{2} = -dt^{2} + (\sinh(\alpha t))^{\frac{2}{n}}\left[\frac{dr^{2}}{1-kr^{2}} + r^{2}\left(d\theta^{2} + 
\sin^{2}\theta d\phi^{2}\right)\right].
\end{equation}
%%%%%%%%%%%%%%%%%%% Figure 4 %%%%%%%%%%%%%%%%%%%%%%%%%%%%%%%%%%%%%%%%%%%%%%%%%%%%%%%%%%%%%%
\begin{figure}[ht]
\centering
\includegraphics[width=8cm,height=8cm,angle=0]{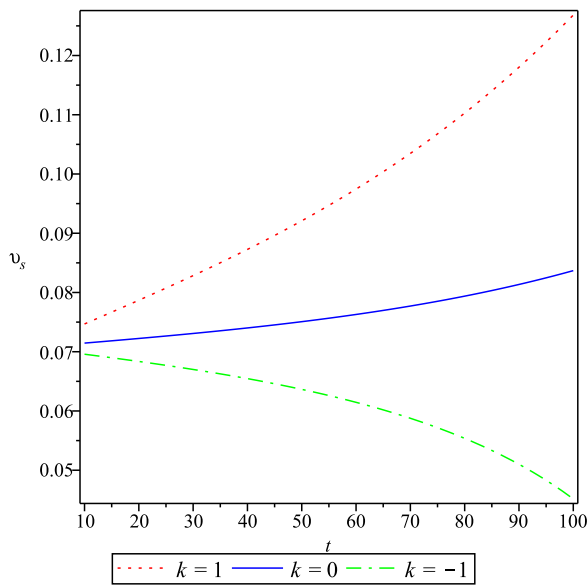} \\
\caption{The plot of sound speed ($\upsilon_{s}$) versus $t$.} 
\end{figure}
%%%%%%%%%%%%%%%%%%%%%%%%%%%%%%%%%% %%%%%%%%%%%%%%%%%%%%%%%%%%%%%%%%%%%%%%%%%%%%%%%%%%%%%%%%%%%
%%%%%%%%%%%%%%%%%%% Figure 5 %%%%%%%%%%%%%%%%%%%%%%%%%%%%%%%%%%%%%%%%%%%%%%%%%%%%%%%%%%%%%%
\begin{figure}[ht]
\centering
\includegraphics[width=8cm,height=8cm,angle=0]{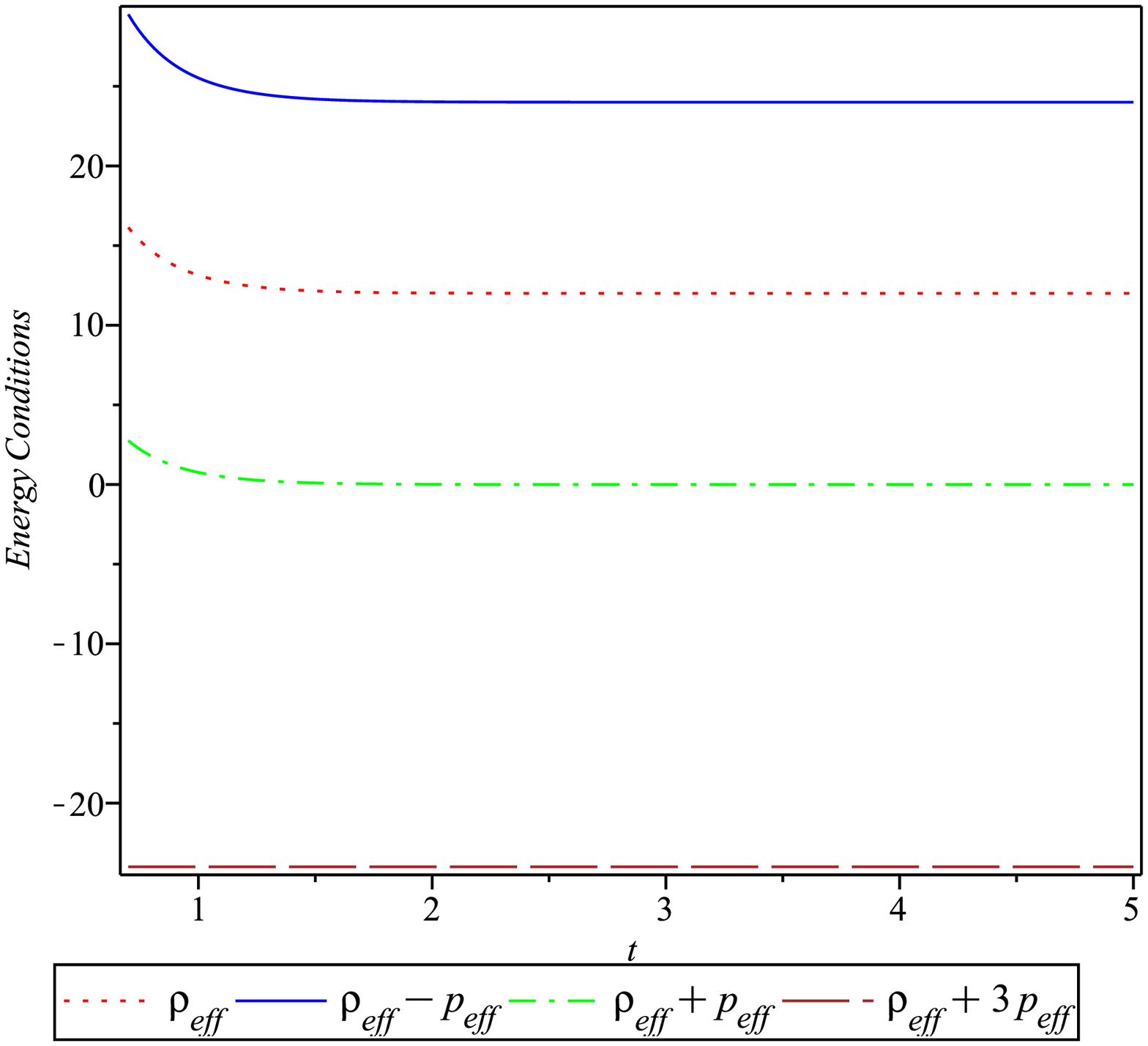} \\
\caption{The plot of energy conditions versus $t$ for open universe ($k = 1$) 
for$\alpha = 2$, $n = 1$.}  
\end{figure}
%%%%%%%%%%%%%%%%%%%%%%%%%%%%%%%%%% %%%%%%%%%%%%%%%%%%%%%%%%%%%%%%%%%%%%%%%%%%%%%%%%%%%%%%%%%%%
%%%%%%%%%%%%%%%%%%% Figure 6 %%%%%%%%%%%%%%%%%%%%%%%%%%%%%%%%%%%%%%%%%%%%%%%%%%%%%%%%%%%%%%
\begin{figure}[ht]
\centering
\includegraphics[width=8cm,height=8cm,angle=0]{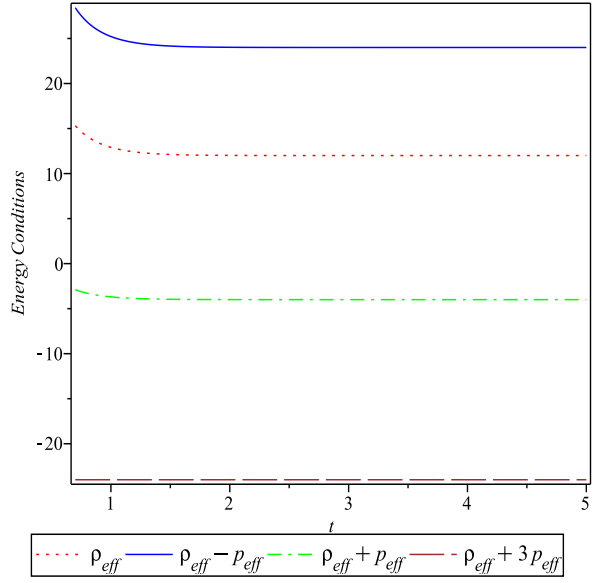} \\
\caption{The plot of energy conditions versus $t$ for flat universe ($k = 0$) 
for $\alpha = 2$, $n = 1$.}  
\end{figure}
%%%%%%%%%%%%%%%%%%%%%%%%%%%%%%%%%% %%%%%%%%%%%%%%%%%%%%%%%%%%%%%%%%%%%%%%%%%%%%%%%%%%%%%%%%%%%
%%%%%%%%%%%%%%%%%%% Figure 5 %%%%%%%%%%%%%%%%%%%%%%%%%%%%%%%%%%%%%%%%%%%%%%%%%%%%%%%%%%%%%%
\begin{figure}[ht]
\centering
\includegraphics[width=8cm,height=8cm,angle=0]{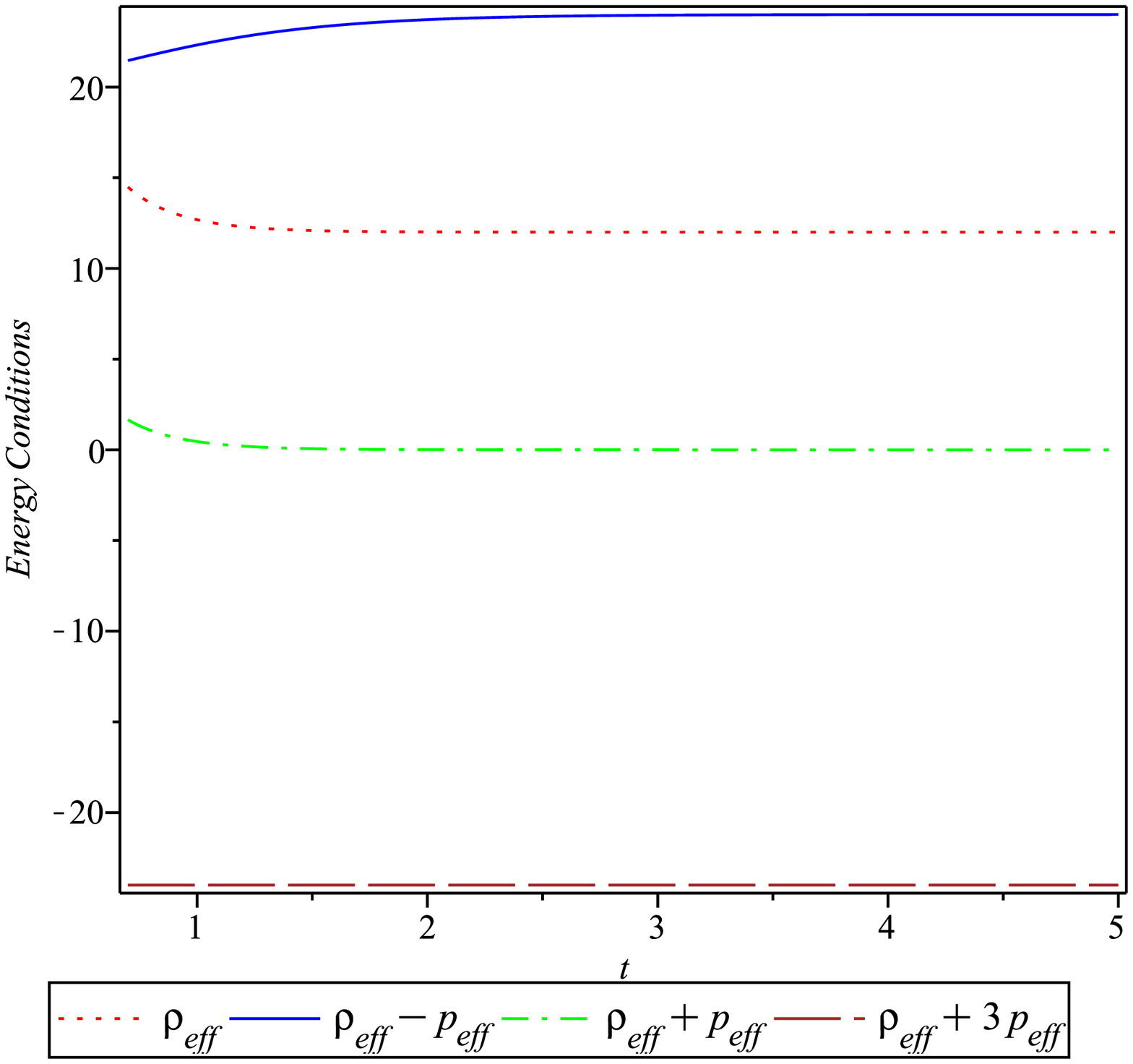} \\
\caption{The plot of energy conditions versus $t$ for closed universe ($k = - 1$) 
for $\alpha = 2$, $n = 1$.}  
\end{figure}
%%%%%%%%%%%%%%%%%%%%%%%%%%%%%%%%%% %%%%%%%%%%%%%%%%%%%%%%%%%%%%%%%%%%%%%%%%%%%%%%%%%%%%%%%%%%%
%%%%%%%%%%%%%%%%%%% Figure 8 %%%%%%%%%%%%%%%%%%%%%%%%%%%%%%%%%%%%%%%%%%%%%%%%%%%%%%%%%%%%%%
\begin{figure}[ht]
\centering
\includegraphics[width=10cm,height=8cm,angle=0]{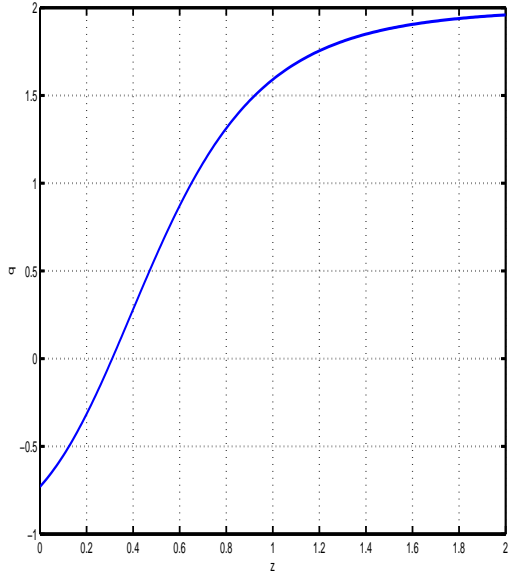} \\
\caption{The plot of deceleration parameter $q$ vs redshift $z$ for $\alpha = 0.137$, $n = 3$.}  
\end{figure}
%%%%%%%%%%%%%%%%%%%%%%%%%%%%%%%%%% %%%%%%%%%%%%%%%%%%%%%%%%%%%%%%%%%%%%%%%%%%%%%%%%%%%%%%%%%%%
%%%%%%%%%%%%%%%%%%%%%%%%%%%%%%%%%%%%%%%%%%%%%%%%%%%%%%%%%%%%%%%%%%%%%%%%%%%%%%%%%%%%%%%%%%%%%%
%%%%%%%%%%%%%%%%%%%%%%%%%%%%%%%  SECTION 4  %%%%%%%%%%%%%%%%%%%%%%%%%%%%%%%%%%%%%%%%%%%%%%%%
\section{Physical and geometric properties of the Model}

The spatial volume ($V$), Hubble parameter ($H$), expansion scalar ($\theta$), energy density ($\rho^{(m)}$) of 
perfect fluid, DE density ($\rho^{(de)}$) and EoS parameter ($\omega^{(de)}$) of DE, for the model (\ref{eq20}) are
found to be
\begin{equation}
\label{eq21} V = (\sinh(\alpha t))^{\frac{3}{n}},
\end{equation}
\begin{equation}
\label{eq22} \theta = 3H = \frac{3\alpha}{n}\coth(\alpha t),
\end{equation}
\begin{equation}
\label{eq23} \rho^{(m)} = \frac{\rho_{0}}{(\sinh(\alpha t))^{\frac{3}{n}(1 + \omega^{(m)})}},
\end{equation}
\begin{equation}
\label{eq24} \rho^{(de)} = \frac{3\alpha^{2}}{n^{2}}(\coth(\alpha t))^{2} + \frac{3k}{(\sinh(\alpha t))^{\frac{2}{n}}} 
- \frac{\rho_{0}}{(\sinh(\alpha t))^{\frac{3}{n}(1 + \omega^{(m)})}},
\end{equation}
\[
\omega^{(de)} = \frac{-1}{\rho^{(de)}}\Biggl[\frac{\alpha^{2}}{n^{2}}(3 - 2n)(\coth(\alpha t))^{2} + 2\frac{\alpha^{2}}{n} + 
\frac{k}{(\sinh(\alpha t))^{\frac{2}{n}}} +
\]
\begin{equation}
\label{eq25} \rho_{0}\frac{\omega^{(m)}}{(\sinh(\alpha t))^{\frac{3}{n}(1+\omega^{(m)})}}
\Biggr].
\end{equation}
The above solutions satisfy energy conservation equations (\ref{eq8}) and (\ref{eq9}) identically. \\

The behavior of EoS for DE ($\omega^{de}$) in term of cosmic time $t$ is shown in Fig. $2$. It is observed that 
for all three closed, flat and open models of the universe, the EoS for DE $\omega^{de}$ is decreasing function 
of time, the rapidity of their falling down at the early stages depend on the type of the universe, while later 
on the EoS parameter for all three models tend to the same constant value $-1$ independent to it. We also observe 
that EoS parameter of closed and flat universe are varying in quintessence era ($\omega^{de} > -1$) through out the 
evolution, while later on they tend to the same constant $-1$ (i.e. cosmological constant) independent to it. We 
also observe that model of open universe started its evolution from quintessence era and crosses the PDL
($\omega^{de} = -1$) and ultimately/finally approaches to $-1$ (i.e. cosmological constant). Therefore, we observe 
that the variation of $\omega^{de}$ in our derived models is in good agreement with recent observations of SNe Ia data 
(Knop et al. 2003), SNe Ia data with CMBR anisotropy and galaxy clustering statistics (Tegmark et al. 2004).\\   

The density parameter $\Omega^{(m)}$ of perfect fluid and density parameter $\Omega^{(de)}$ of DE are given by
\begin{equation}
\label{eq26} \Omega^{(m)} = \frac{\rho^{(m)}}{3H^{2}} = \frac{\rho_{0}n^{2}}{3\alpha^{2}}\frac{(\sinh(\alpha t))^{(2 - \frac{3}{n}
(1 + \omega^{(m)}))}}{(\cosh(\alpha t))^{2}},
\end{equation}
\begin{equation}
\label{eq27} \Omega^{(de)} = \frac{\rho^{(de)}}{3H^{2}} = 1 + \frac{n^{2}}{3\alpha^{2}}(\tanh(\alpha t))^{2} 
\left[\frac{k}{(\sinh(\alpha t))^{\frac{2}{n}}} - \frac{\rho_{0}}{(\sinh(\alpha t))^{\frac{3}{n}(1 + \omega^{(m)})}}\right].
\end{equation}
Adding (\ref{eq26}) and (\ref{eq27}), we get the overall density parameter
\begin{equation}
\label{eq28} \Omega = \Omega^{(m)} + \Omega^{(de)} = 1 + \frac{k}{3}\left(\frac{n\tanh(\alpha t)}{\alpha (\sinh(\alpha t))^
{\frac{1}{n}}}\right)^{2}.
\end{equation}
From the right hand side of Eq. (\ref{eq28}) it is clear that in flat universe $(k = 0)$, $\Omega = 1$ and in open universe 
$(k = -1)$, $\Omega < 1$ and in closed universe $(k = +1)$, $\Omega > 1$. But at late time we see for all flat, open and 
closed universes $\Omega \to 1$. This result is also compatible with the observational results. Since our model predicts 
a flat universe for large times and the present-day universe is very close to flat, so the derived model is also compatible 
with the observational results (Bennett et al. 2003; Tegmark et al. 2004a). The variation of density parameter with cosmic 
time has been shown in Fig. $3$. \\
%%%%%%%%%%%%%%%%%%%%%%%%%%%%%%%%%%%%%%%%%%%%%%%%%%%%%%%%%%%%%%%%%%%%%%%%%%%%%%%%%%%%%%%%%%%%%%%%%%%
%%%%%%%%%%%%%%%%%%%%%%%%%%%%%%%% SECTION 5  %%%%%%%%%%%%%%%%%%%%%%%%%%%%%%%%%%%%%%%%%%%%%%%%%%%%%%%%%%%
\section{Physical acceptability of solution}
For the stability of corresponding solution, we should check that our model is physically acceptable. For this, firstly it 
is required that the velocity of sound should be less than velocity of light i.e. within the range $0 \leq \upsilon_{s} = 
\left(\frac{dp^{de}}{d\rho^{de}}\right) \leq 1$. \\

The sound speed is obtained as:
\begin{equation}
\label{eq29}
\upsilon_{s} = \sqrt{\frac{(2n -3)K_{1} + K_{2} + 3 \omega^{m} K_{3}}{3(K_{1} - K_{2} + K_{3})}},
\end{equation}
where
$$
K_{1} = -\frac{\alpha^{3}}{n^{2}}\coth(\alpha t) csch^{2}(\alpha t),
$$
$$
K_{2} = \frac{\alpha k}{n}\left(\cosh(\alpha t)\right)\left[\sinh(\alpha t)\right]^{\left(\frac{-2}{n} - 1\right)},
$$
$$
K_{3} = \frac{\alpha \rho_0}{2n} \left(1 + \omega^{m}\right)\left(\cosh(\alpha t)\right)\left[\sinh(\alpha t)\right]^{\left(\frac{-3}{n}
\left(1 + \omega^{m}\right) - 1\right)}.
$$
In this cases we observe that $\upsilon_{s} < 1$. From Fig. $4$ depicts the plot of sound speed ($\upsilon_{s}$) versus cosmic time $t$.
It is sort out from the figure that speed of sound remains less than the speed of light ($c = 1$) throughout the evolution of the universe. \\

Secondly, the weak energy conditions (WEC) and dominant energy conditions (DEC) are given by \\

(i) $\rho_{eff} \geq 0$, ~ ~ (ii) $\rho_{eff} - p_{eff} \geq 0$ ~ ~ and ~ ~ (iii) $\rho_{eff} + p_{eff} \geq 0$. \\

The strong energy conditions (SEC) are given by $\rho_{eff} + 3p_{eff} \geq 0$. \\

From straight forward calculation, we obtain the following expression:
\begin{flushleft}
(i) $\rho_{eff}\geq0$ implies that\\
\end{flushleft}
\begin{equation}
\label{eq30}
\frac{3\alpha^{2}}{n^{2}}\left(\coth\left(\alpha t\right)\right)^{2} + 3k\left(\sinh(\alpha t)\right)^{(\frac{-2}{n})} \geq 0.
\end{equation}
\begin{flushleft}
(ii) $\rho_{eff}-p_{eff}\geq0$ implies that\\
\end{flushleft}
\begin{equation}
\label{eq31}
\frac{\alpha^{2}}{n^{2}}\left(6-2n\right)\left(\coth\left(\alpha t\right)\right)^{2} + \frac{2\alpha^2}{n} + 4k\left(\sinh(\alpha t)
\right)^{(\frac{-2}{n})} \geq 0.
\end{equation}
\begin{flushleft}
(iii) $\rho_{eff}+p_{eff} \geq0$ implies that\\
\end{flushleft}
\begin{equation}
\label{eq32}
\frac{2\alpha^2}{n}\left(\coth\left(\alpha t\right)\right)^2-\frac{2\alpha^2}{n}+2k\left(\sinh(\alpha t)\right)^{(\frac{-2}{n})}\geq0.
\end{equation}
\begin{flushleft}
(iv) $\rho_{eff}+3p_{eff}\geq0$ implies that \\
\end{flushleft}
\begin{equation}
\label{eq33}
\frac{6\alpha^2}{n^2}\left(-1+n\right)\left(\coth\left(\alpha t\right)\right)-\frac{6\alpha^2}{n}\geq0.
\end{equation}
From the Figures $5$ $-$ $7$, we observe that 
\begin{itemize}
\item The WEC and DEC for both the open and closed universes are satisfied but SEC is violated as expected.  
\end{itemize}
\begin{itemize}
\item In flat model, the WEC is satisfied whereas the DEC and SEC are violated through the whole evolution of the universe.
\end{itemize}
Therefore, on the basis of above discussions and analysis, our corresponding solutions are physically acceptable. \\ 
%%%%%%%%%%%%%%%%%%%%%%%%%%%%%%%  SECTION 6  %%%%%%%%%%%%%%%%%%%%%%%%%%%%%%%%%%%%%%%%%%%%%%%%
\section{Conclusions}
In this present work we continue and extend the previous work of Amirhashchi et al. (2011b) and Saha et al. (2012).
In summary, we have studied a system of two fluid within the scope of a spatially homogeneous and isotropic 
FRW model. The role of two fluid minimally coupled in the evolution of the dark energy parameter has been 
investigated. The field equations have been solved exactly with suitable physical assumptions. The solutions satisfy 
the energy conservation equation identically. Therefore, exact and physically viable FRW model has been obtained.
It is to be noted that our method of solving the field equations is different from the technique of Kumar (2011).
Kumar has solved the field equations by considering the constant DP whereas we have considered time-dependent DP. 
As we have already mentioned in previous section that for a universe which was decelerating in past and accelerating 
at the current epoch, the DP must show signature flipping (Padmanabhan and Roychowdhury 2003; Amendola 2003; Riess 
et al. 2001). So, it is reasonable to consider time dependent DP. The main features of the model are as follows:

\begin{itemize}
\item The present DE model has a transition of the universe from the early deceleration phase to the recent acceleration 
phase (see, Figure $1$) which is in good agreement with recent observations (Caldwell et al. 2006). 

\item The DP ($q$) as a function of the red shift parameter $z = -1 + \frac{a_0}{a}$, where $a_{0}$ is the
present value of the scale factor i.e. at $z=0$, is given by
\begin{equation}
\label{eq34} q(z) = n - 1 - n\left[\tanh\left(\sinh^{-1}\sqrt{\frac{n-1-q_{0}}{(q_{0}+1)(z+1)
^{2n}}}\right)\right]^{2}.
\end{equation}
Here $q_{0}$ is the present value of deceleration parameter i.e. at $z = 0$. If we set $q_{0} = -0.73$ (Cunha and Lima 2008)
for the present Universe ($t_{0} = 13.7$ GYr), we get the following relationship between the constants $n$ and $\alpha$:
\begin{equation}
\label{eq35} \alpha = \frac{1}{13.7}\tanh^{-1}\left[1-\frac{0.27}{n}\right]
^{\frac{1}{2}}.
\end{equation}
It is self explanatory from the above relation that for the present universe, the model is valid only for $n > 0.27$.
Figure $4$ depicts the behaviour of $q$ with red shift $z$ for $q_{0} = -0.73$ and for a representative case, we have
chosen $n = 3$ and $\alpha = 0.137$ satisfying the above Eq. (\ref{eq35}). It is clearly observable from the Fig. $8$
that the transition red shift ($z_{t}$) for our model with $q_{0} = -0.73$  is $\cong 0.32$. This is in good agreement 
with the cosmological observations in the literature (Cunha and Lima 2008; Cunha 2009; Pandolfi 2009; Lima et al. 2010; 
Li et al. 2011), according to which the transition red shift ($z_{t}$) of the accelerating expansion is given by 
$0.3 < z_{t} < 0.8$. In particular, the kinematic approach to cosmological data analysis provides a direct evidence to 
the present accelerating stage of the universe, which does not depend on the validity of general relativity, a well as 
on the matter-energy content of the universe (Cunha and Lima 2008).

\item It is observed that EoS parameter of closed and flat universe are varying in quintessence era ($\omega^{de} > -1$) 
through out the evolution, while later on they tend to the same constant $-1$ (i.e. cosmological constant) independent to it. 

\item It is observed that the open universe started its evolution from quintessence era and crosses the PDL
($\omega^{de} = -1$) and finally approaches to $-1$ (i.e. cosmological constant). Thus, we find that the EoS parameter 
for open universe changes from $\omega^{de} > -1$ to $\omega^{de} < -1$, which is consistent with recent observations. 

\item The total density parameter ($\Omega$) approaches to $1$ for sufficiently large time (see, Figure $3$) which 
is reproducible with current observations. 

\item For different choice of $n$, we can generate a class of DE models in FRW universe. It is observed that such DE 
models are also in good harmony with current observations. For example, if we put $n = 1$ in the present paper, we 
obtain all results of recent paper of Amirhashchi et al. (2011b). \\ 

\item Thus, the solutions demonstrated in this paper may be useful for better understanding of the characteristic of 
DE in the evolution of the universe within the framework of FRW space-time.

\end{itemize}
%%%%%%%%%%%%%%%%%%%%%%%%%%%%%%%%%%%%%%%%%%%%%%%%%%%%%%%%%%%%%%%%%%%%%%%%%%%%%%%%%%%%%%%%%%%%%%%%%%%%%%%%%%
\section*{Acknowledgments}
Author (A. Pradhan) would like to thank the Inter-University Centre for Astronomy and Astrophysics (IUCAA), 
Pune, India for providing facility and support where part of this work was carried out. This work was 
supported by the University Grants Commission, New Delhi, India under the grant (Project  F.No. 41-899/2012 (SR)).
The author also thank Prof. H. Amirhashchi for his fruitful suggestions.
%%%%%%%%%%%%%%%%%%%%%%%%%%%%%%%%%%%%%%%%%%%%%%%%%%%%%%%%%%%%%%%%%%%%%%%%%%%%%%%%%


\begin{thebibliography}{99}
\bibitem {ref1}
Akarsu,$\ddot{O}$., Kilinc, C.B.: Gen. Relat. Gravit. {\bf 42} 119 (2010a)
\bibitem {ref2}
Akarsu, $\ddot{O}$., Kilinc, C.B.: Gen. Relat. Gravit. {\bf 42} 763 (2010b)
\bibitem {ref3}
Akarsu, $\ddot{O}$., Kilinc, C.B.: Astrophys. Space Sci. {\bf 326} 315 (2010c)
\bibitem {ref4}
Akarsu,$\ddot{O}$., Dereli, T.: Int. J. Theor. Phys. {\bf 51}, 612 (2012)
\bibitem {ref5}
Amirhashchi, H., Pradhan, A., Zainuddin, H.: Res. Astron. Astrophys. {\bf 13}, 129 (2013)
\bibitem {ref6}
Amendola, L.: Mon. Not. R. Astron. Soc. {\bf 342}, 221 (2003)
\bibitem {ref7}
Amirhashchi, H., Pradhan, A., Saha, B.: Chin. Phys. Lett. {\bf 28}, 039801 (2011a)
\bibitem {ref8}
Amirhashchi, H., Pradhan, A., Zainuddin, H.: Int. J. Theor. Phys. {\bf 50}, 3529 (2011b)
\bibitem {ref9}
Astier, P., et al.: Astron. Astrophys. {\bf 447}, 31 (2006)
\bibitem {ref10}
Bennett, C.L., et al.: Astrophys. J. Suppl. {\bf 148}, 1 (2003)
\bibitem {ref11}
Cai, Y.-F., Saridakis, E.N., Xia, J.-Q.: Phys. Rep. {\bf 493}, 1 (2010). hep-th/0909.2776
\bibitem {ref12}
Caldwell, R.R., Komp, W., Parker, L., Vanzella, D.A.T.: Phys. Rev. D {\bf 73}, 023513 (2006)
\bibitem {ref13}
Carroll, S.M., Hoffman, M.: Phys. Rev. D. {\bf 68}, 023509 (2003)
\bibitem {ref14}
Chawla, C., Mishra, R.K.: Rom. J. Phys. {\bf 58}, 75 (2013)
\bibitem {ref15}
Chawla, C., Mishra, R.K., Pradhan, A.: Eur. Phys. J. Plus {\bf 127}, 137 (2012) 
\bibitem {ref16}
Chen, X., Gong, Y., Saridakis, E.N.: JCAP {\bf 0904}, 001 (2009). 0812.1117/gr-qc.
\bibitem {ref17}
Cunha, J.V., Phys. Rev. D {\bf 79}, 047301 (2009)
\bibitem {ref18}
Cunha, J.V., Lima, J.A.S.: Mon. Not. R. Astron. Soc. {\bf 390}, 210 (2008)
\bibitem {ref19}
Clocchiatti, A., et al. (High Z SN Search Collaboration): Astrophys. J. {\bf 642}, 1 (2006)
\bibitem {ref20}
Copeland, E., Sami, M., Tsujikawa, S.: Int. J. Mod. Phys. D {\bf 15}, 1753 (2006). arXiv:0603057[hep-th]
\bibitem {ref21}
de Bernardis, P., et al.: Nature {\bf 666}, 716 (2007)
\bibitem {ref22}
Davis, T.M., et al., Astrophys. J. {\bf 598}, 102 (2003)
\bibitem {ref23}
Eisentein, D.J., et al.: Astrophys. J. {\bf 633}, 560 (2005)
\bibitem {ref24}
Hanany, S., et al.: Astrophys. J. {\bf 545}, L5 (2000)
\bibitem {ref25}
Hinshaw, G., et al.: Astrophys. J. Suppl. {\bf 180}, 225 (2009)
\bibitem {ref26}
Jamil, M., Saridakis, E.N., Setare, M.R.: Phys. Rev. D {\bf 81}, 023007 (2010). hep-th/0910.0822
\bibitem {ref27}
Knop, R.K., et al.: Astrophys. J. {\bf 598}, 102 (2003)
\bibitem {ref28}
Komatsu, E., et al.: Astrophys. J. Suppl. Ser. {\bf 180}, 330 (2009)
\bibitem {ref29}
Kumar, S.: Astrophys. Space sci. {\bf 332}, 449 (2011)
\bibitem {ref30}
Li, Z., Wu, P., Yu, H.: Phys. Lett. B {\bf 695}, 1 (2011) 
\bibitem {ref31}
Lima, J.A.S., Holanda, R.F.L., Cunha, J.V.: AIP Conf. Proc. {\bf 1241}, 224 (2010)
\bibitem {ref32}
Kumar, S., Yadav, A.K.: Mod. Phys. Lett. A {\bf 26}, 647 (2011)
\bibitem {ref33}
MacTavish, C.J., et al.: Astrophys. J. {\bf 647}, 799 (2006)
\bibitem {ref34}
Mishra, R.K., Pradhan, A., Chawla, C.: Int. J. Theor. Phys. DOI 10.1007/s10773-013-1540-4 (2013)
\bibitem {ref35}
Naidu, R.L., Satyanarayana, B., Reddy, D.R.K.: Int. J. Theor. Phys. {\bf 51}, 1997 (2012a)
\bibitem {ref36}
Naidu, R.L., Satyanarayana, B., Reddy, D.R.K.: Int. J. Theor. Phys. {\bf 51}, 2857 (2012b)
\bibitem {ref37}
Padmanabhan, T., Roychowdhury, T.: Mon. Not. R. Astron. Soc. {\bf 344}, 823 (2003)
\bibitem {ref38}
Pandolfi, S.: Nucl. Phys. B {\bf 194}, 294 (2009)
\bibitem {ref39}
Perlmutter, S., et al. (Supernova Cosmology Project Collaboration): Nature {\bf 391}, 51 (1998)
\bibitem {ref40}
Perlmutter, S., et al. (Supernova Cosmology Project Collaboration): Astrophys. J. {\bf 517}, 5 (1999)
\bibitem {ref41}
Pradhan, A., Otarod, S.: Astrophys. Space Sci. {\bf 306}, 11 (2006)
\bibitem {ref42}
Pradhan, A., Amirhashchi, H., Saha, B.: Astrophys. Space Sci. {\bf 333}, 343 (2011)
\bibitem {ref43}
Pradhan, A., Jaiswal, R., Jotania, K., Khare, R.K.: Astrophys. Space Sci. {\bf 337}, 401 (2012)
\bibitem {ref44}
Pradhan, A.: Res. Astron. Astrophys. {\bf 13}, 139 (2013)
\bibitem {ref45}
Pradhan, A., Singh, A.K., Chouhan, D.S.: Int. J. Theor. Phys. {\bf 52}, 266 (2013)
\bibitem {ref46}
Pradhan, A., Jaiswal, R., Khare, R.K.: Astrophys. Space Sci. {\bf 343}, 489 (2013)
\bibitem {ref47}
Reddy, D.R.K., Kumar, R.S.: Int. J. Theor. Phys. {\bf 52}, 1362 (2013)
\bibitem {ref48}
Reddy, D.R.K., Satyanarayana, B., Naidu, R.L.: Astrophys. Space Sci. {\bf 339}, 401 (2012)
\bibitem {ref49}
Riess, A.G., et al. (Supernova Search Team Collaboration): Astron. J. {\bf 116}, 1009 (1998)
\bibitem {ref50}
Riess, A.G., et al.: Astrophys. J. {\bf 659}, 98 (2007)
\bibitem {ref51}
Riess, A.G., et al.: Astrophys. J. {\bf 560}, 49 (2001)
\bibitem {ref52}
Riess, A.G., et al. (Supernova Search Team Collaboration): Astrophys. J. {\bf 607}, 665 (2004)
\bibitem {ref53}
Saha, B., Amirhashch, H., Pradhan, A.: Astrophys. Space Sci. {\bf 342}, 257 (2012)
\bibitem {ref54}
Singh, T., Chaubey, R.: Res. Astron. Astrophys. {\bf 12}, 473 (2012)
\bibitem {ref55}
Singh, T., Chaubey, R.: Canad. J. Phys. {\bf 91}, 180 (2013)
\bibitem {ref56}
Tegmark, M., et al. (SDSS collaboration): Phys. Rev. D {\bf 69}, 103501 (2004)
\bibitem {ref57}
Tonry, J.L., et al. (Supernova Search Team Collaboration): Astrophys. J. {\bf 594}, 1 (2003)
\end{thebibliography}
\end{document}